\documentclass[aps,preprint,a4paper,showpacs,amsmath,amssymb,floatfix,prb]{revtex4}
\usepackage{graphicx}

\begin{document}

\title{Qualitative modifications and new dynamic phases in the phase diagram of 1D superconducting wires driven with electric currents}
\author{Shimshon Kallush and Jorge Berger}
\affiliation{Department of Physics and Optical Engineering, ORT-Braude College, P.O. Box 78, 21982 Karmiel, Israel}

\begin{abstract}
After an initial transient period, the conduction regime in a 1D superconducting wire that carries a fixed current is either normal or periodic or stationary. The phase diagram for these possibilities was studied in Phys. Rev. Lett. {\bf 99}, 167003 (2007) for particular values of the length and the material parameters. We have extended this study to arbitrary length and to a range of material parameters that includes realistic values. Variation of the length leads to scaling laws for the phase diagram. Variation of the material parameters leads to new qualitative features and new phases, including a parameter region in which all three regimes are possible. 
\end{abstract}

\pacs{74.25.Dw, 74.20.De, 74.81.Fa, 74.25.Sv}

\maketitle

\section{Introduction}
One of the classic problems in superconductivity is that of pattern formation in current carrying filaments, which has remained an active field since decades ago\cite{AL,Beasley,Bara,Ivlev,Tidecks} to present.\cite{suppress,S,koby,Lydia} 
An insightful procedure to obtain the phase diagram for this system, which has found several features that had been previously ignored, was advanced in Ref.~\onlinecite{koby}.
The theoretical tools that have been applied to this problem are the simple time-dependent Ginzburg--Landau model\cite{TDGL}(TDGL) and the Kramer--Watts-Tobin model.\cite{KWT} Although for realistic situations there are large quantitative differences between the results predicted by these two models, it was found\cite{koby} that the qualitative behavior, on which we intend to focus, is the same for both. 

Theoretical studies have found\cite{Beasley,Ivlev,Tidecks,S,koby} that, after an initial transient time, the order parameter function (and thus the current pattern) stabilizes at one of four regimes: (i) normal (N), in which there is no superconducting condensate; (ii) stationary (S), in which superconducting and normal currents flow in parallel and do not depend on time; (iii) periodic (P), in which the superconducting current and all measurable quantities are periodic functions of time; and (iv) fully superconducting, in which there is no normal current. Following references \onlinecite{Bara} and \onlinecite{koby}, we will consider the situation in which the contacts are normal metals and require that the order parameter vanishes at the boundaries, thus ruling out the fully superconductive regime.

A salient feature of the periodic regime is the existence of phase slip centers, at which the superconducting order parameter vanishes periodically in time. In the parameter region that we will consider (short wires and temperatures close to $T_c$), there will be at most one phase slip center.

\begin{figure}[tbp] 
 \centerline{
    \mbox{\includegraphics[width=3.00in]{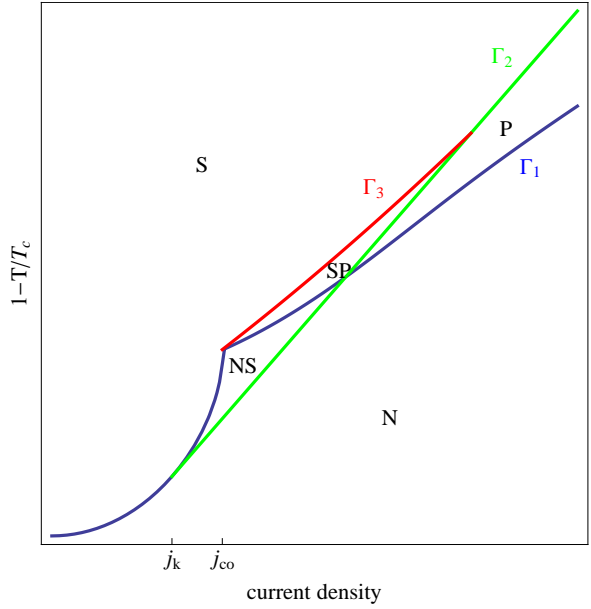}}}
\caption{(Color online) Phase diagram found with the parameters adopted in Ref.~\onlinecite{koby}. N (S, P) marks the region where the normal (respectively, stationary, periodic) regime exists and is stable. There are small regions where two regimes are stable. The temperature range in this diagram is below (and does not include) $T_c$, but is limited to a region close to $T_c$.} 
\label{qual}
\end{figure}

In the high temperature and low current region, the findings of Ref.~\onlinecite{koby} may be qualitatively summarized as in Fig.~\ref{qual}. In this diagram there are three critical lines: (i) above the line $\Gamma_1$ the normal regime N is unstable (this line has a Hopf singularity at current density $j_{\rm co}$); (ii) below the line $\Gamma_2$ the stationary regime S is unstable; and (iii) at $\Gamma_3$ the periodic regime P degenerates into S (for high currents, $\Gamma_2$ and $\Gamma_3$ coalesce). Likewise, P degenerates into N at $\Gamma_1$. Accordingly, there are regions in which N, S or P are stable, including small areas where two regimes are stable.

In Ref.~\onlinecite{koby}, particular values of the geometric and material parameters were adopted. In this article we show that variation of these parameters leads to qualitative changes in the phase diagram. 

\section{Model}
Following Ref.~\onlinecite{koby}, we use the time-dependent Ginzburg-Landau model, choose a gauge with no vector potential, write $\Gamma =1-T/T_c$  and  denote by $\varphi$ the electrochemical potential. The unit of length will be denoted by $x_0$ , the unit of time by $t_0$ , the unit of voltage by $\varphi_0$, and the unit of the current density  by $j_0$. We take
\begin{equation}
x_0=\xi(0)\;,\;\;t_0=\frac{\pi\hbar }{8k_BT_c}\;,\;\;\varphi_0=\frac{4k_BT_c}{\pi e}\;,\;\;
j_0=\frac{4\sigma  k_BT_c}{\pi e\xi (0)}\;.
\label{units}
\end{equation}
Here $\xi (0)$ is the coherence length at $T=0$, $k_B$ is Boltzmann's constant, $e$ is the electron charge, and $\sigma $ is the normal conductivity.

With this notation, the TDGL equation and Ohm's law read
\begin{equation}
\psi _t  = \psi _{xx}  + \left( {\Gamma  -  \left| \psi  \right|^2  - i\varphi } \right)\psi \;
\label{tdgl}
\end{equation}
and
\begin{equation}
\varphi _x  = u\, {\rm Im}\left( { \psi _x \psi ^* } \right) - j\;.
\label{Ohm}
\end{equation}
Here $\psi$ is the order parameter, with normalization imposed by Eq.~(\ref{tdgl}), the subscripts denote partial differentiation with respect to the time $t$ and the arc length $x$ along the wire, and $u$ is the ratio between the relaxation times of $\psi$ and $j$.\cite{Kopnin} 

The wire is assumed to extend along $-L\le x\le L$. We take the origin of the potential at the middle of the wire, $\varphi (0)=0$, and assume that the contacts at the end of the wire are made of a normal metal, so that the boundary conditions are $\psi (\pm L)=0$. In Ref.~\onlinecite{koby} the values of $L$ and $u$ were set as 1. The value $u=1$ corresponds to a clean superconductor in which the the electron mean free path is about five times longer than the BCS coherence length.\cite{Kopnin}

Since for realistic situations $L\gg x_0$ and since the most usual situation for which TDGL is applicable is that of a dirty superconductor with paramagnetic impurities, the feasible values of $u$ and $L$ may vary from unity. More realizable values are  $u\sim 5$ and $L \sim 10^2$. We will thus investigate how the phase diagram in Fig.~\ref{qual} changes when $L$ and $u$ vary.

We note that $\Gamma_1$ is a bifurcation from the normal phase $\psi =0$, so that the term containing $\psi _x \psi ^*$ in Eq.~(\ref{Ohm}) vanishes at $\Gamma_1$. As a consequence, the line $\Gamma_1(j)$ is independent of $u$, and so is its Hopf singularity $j_{co}$.

\section{Procedures and Results}
\subsection{Dependence on $L$\label{LL}}
We will present now the scaling laws with respect to $L$. Inspection of Eqs. (\ref{tdgl}) and (\ref{Ohm}) reveals that for every solution of these equations with boundary conditions $\psi (\pm 1)=0$, a corresponding solution with boundary conditions $\psi (\pm L)=0$ can be automatically obtained. In order to fulfill the new boundary condition, one should transform $x\rightarrow Lx$. Additionally, the complementary transformations $t\rightarrow L^2t$, $\psi\rightarrow L^{-1}\psi$, $\varphi \longrightarrow L^{-2}\varphi $, $\Gamma \rightarrow L^{-2}\Gamma $ and $j\rightarrow L^{-3}j$ should be carried. By doing this, each of the terms in Eqs. (\ref{tdgl}) and (\ref{Ohm}) will be multiplied by $L^{-3}$, so that for any given solution, the transformed order parameter function will be a solution of the transformed equations. In particular, the phase diagram in Fig.~\ref{qual} will not change qualitatively, but will be scaled: the current density will be scaled by a factor of $L^{-3}$ and $1-T/T_c$ will be scaled by a factor of $L^{-2}$.

\subsection{Numerical Methods}
Equations (\ref{tdgl}) and (\ref{Ohm}) were solved numerically. 
Spatial derivatives of the order parameter may be obtained by simple finite differences, but doing so we obtained values of $\psi_{xx}$ that were too inaccurate near the boundaries, and eventually lead to instability. Therefore, spatial derivatives were evaluated by the Fourier method:
$\psi$ is transformed to Fourier space, multiplied by the wave vector $ik$ (where $k$ is the reciprocal variable of $x$) and transformed back to position space.
For more detailed description see \cite{k104}. Temporal integration was taken by a simple first order finite difference (Euler iterations), after verifying the convergence with higher approximation order for the time derivative. To determine the character of the asymptotic behavior at a given point $(\Gamma ,j)$ in the phase diagram, several initial values for $\psi (x)$ were taken, and then propagated until a stationary or periodic regime was reached.
The final functions that were obtained served as the initial values for the nearby points in the phase diagram.

\begin{figure}[tbp] 
 \centerline{
    \mbox{\includegraphics[width=3.00in]{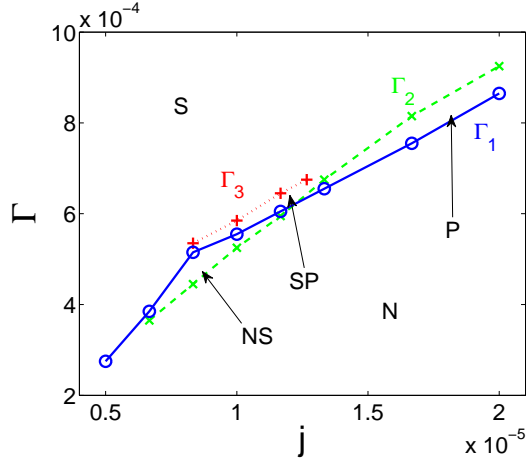}}}
\caption{(Color online) Typical phase diagram for $u\approx 1$. Symbols denote computed phase transition points: blue circles confine the normal phase from above, green ``x" markers confine the stationary phase from below, and red crosses encapsulate the SP phase. Wire parameters: $L = 120$, $u=1.67$.} 
\label{fig3}
\end{figure}

For values of $u$ close to 1 we recovered the three transition lines in Fig.~\ref{qual}. Figure \ref{fig3} shows a typical result. There is a transition line $\Gamma_1(j)$, such that the N and NS phases are located below it and there is no stable normal phase for $\Gamma > \Gamma_1(j)$. 
$\Gamma_2(j)$ limits the NS and the S phases from below. For $\Gamma > \Gamma_2(j)$ there is either a stable or a metastable S phase. Finally, $\Gamma_3(j)$ connects the two other lines and is the upper bound for the SP phase. 

As a test of our numerical procedure, we checked the scaling $\Gamma_i(j;L')=(L/L')^2\Gamma_i(L^3j/L'^3;L)$, as predicted in Sec.~\ref{LL}, for $i=1,2$ and $60\le L,L'\le 480$, and found perfect agreement.

\begin{figure}[tbp] 
 \centerline{
    \mbox{\includegraphics[width=3.00in]{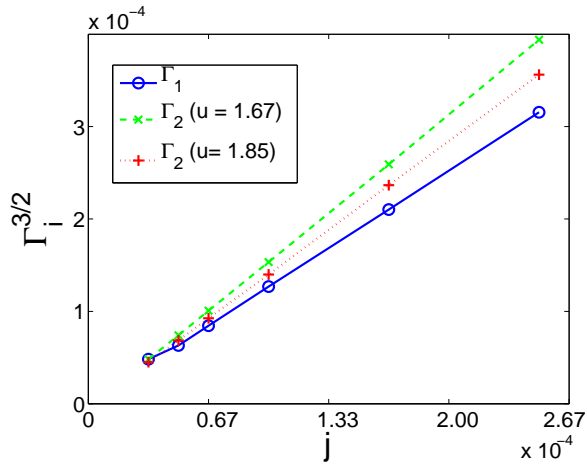}}}
\caption{(Color online) Asymptotes of the curves $\Gamma_1^{3/2}$ and $\Gamma_2^{3/2}$ for two values of $u$ and large current densities. For $j\agt 6\times 10^{-5}$ the lines are straight within the numerical accuracy. $L=120$. } 
\label{fig2a}
\end{figure}

We investigated the asymptotic behavior of $\Gamma_1(j)$ and $\Gamma_2(j)$ for large current densities. In Ref.~\onlinecite{Zum} it was shown that the leading term of $\Gamma_1$ obeys $\Gamma_1 \propto j^{2/3}$. Motivated by this result, in Fig.~\ref{fig2a} we plot $\Gamma_1(j)^{3/2}$ and $\Gamma_2(j)^{3/2}$ for several values of $u$. As could be expected, $\Gamma_1(j)^{3/2}$ is well described by a linear function for $j\agt 6\times 10^{-5}$.  It turns out that in this region $\Gamma_2(j)^{3/2}$ is also well described by a linear function.

\subsection{Dependence on $u$ and phase diagram modifications}
For $u = L = 1$ the phase diagram looks as in Fig.~\ref{qual}, and Fig.~\ref{fig3} shows that this is still the case for $u=1.67$. However, as shown in Fig.~\ref{fig2a}, increasing $u$ results in a lower slope of $\Gamma_2^{3/2}$, so that for sufficiently large $u$ we expect that $\Gamma_2$ will not intersect with $\Gamma_1$, thus leading to a qualitatively different phase diagram.

\begin{figure}[tbp] 
 \centerline{
    \mbox{\includegraphics[width=3.00in]{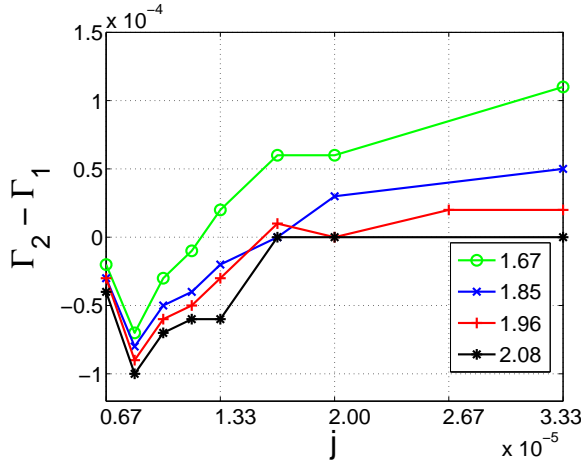}}}
\caption{(Color online) Position of $\Gamma_2$ relative to $\Gamma_1$: $\Gamma_2-\Gamma_1$ for different values of $u$, as indicated in the legend.} 
\label{fig4}
\end{figure}

We have therefore evaluated $\Gamma_1(j)$ and $\Gamma_2(j)$ for $u\agt 1.67$. Our results are presented in Fig.~\ref{fig4}. We want to focus on the NS, SP and P regions of the diagram, which describe the qualitative stages in the overall transition between N and S. Since for large values of $u$ the various curves are in close proximity and some of the phases span very narrow regions, we enlarge their visibility by plotting $\Gamma_2(j)-\Gamma_1(j)$. In all the figures from now on we will present the results for $L=120$. The regions where $\Gamma_2(j)<\Gamma_1(j)$ imply the presence of the NS phase, whereas the regions where $\Gamma_2(j)>\Gamma_1(j)$ imply the presence of the P phase. As suspected, Fig.~\ref{fig4} suggests that $\Gamma_2(j)$ is never larger than $\Gamma_1(j)$ for $u = 2.08$.

\begin{figure}[tbp] 
 \centerline{
    \mbox{\includegraphics[width=3.00in]{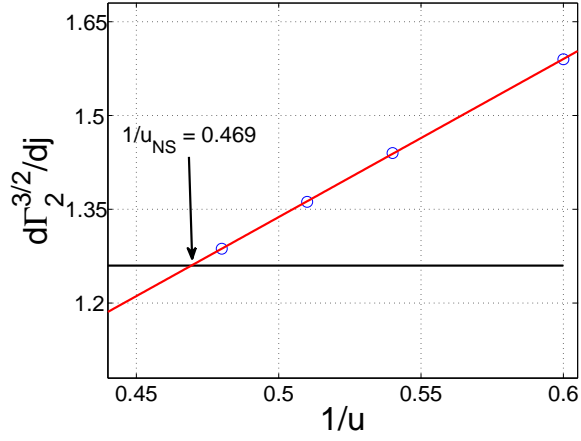}}}
\caption{(Color online) Determination of the critical value $u=u_{\rm NS}$. $\Gamma_1(j)$ and $\Gamma_2(j)$ intersect if and only if the asymptotic slope of $\Gamma_2^{3/2}$ is larger than that of $\Gamma_1^{3/2}$, $d\Gamma_1^{3/2}/dj=1.26$ which is the case for $u<u_{\rm NS}$. The slopes have been plotted as functions of $1/u$ in order to take advantage of the apparently linear behavior.} 
\label{fig5}
\end{figure}

We will now determine the critical value of $u$, $u_{\rm NS}$, at which $\Gamma_1(j)$ and $\Gamma_2(j)$ cease intersecting each other. Our results indicate that for small currents $\Gamma_2(j)<\Gamma_1(j)$. As long as the slope of $\Gamma_2(j)$ is larger than that of $\Gamma_1(j)$, there will be a value of $j$ where $\Gamma_2(j)=\Gamma_1(j)$. However, Fig.~\ref{fig2a} indicates that the slope of $\Gamma_2(j)$ decreases as $u$ increases, and the current for which $\Gamma_2(j)=\Gamma_1(j)$ will increase accordingly, until the lines no longer intersect. Figure \ref{fig5} displays the asymptotic slopes of $\Gamma_2^{3/2}(j)$ and $\Gamma_1^{3/2}(j)$ as a function of $1/u$. A reasonable extrapolation indicates that $u_{\rm NS}=2.13$.

\begin{figure}[tbp] 
 \centerline{
    \mbox{\includegraphics[width=3.00in]{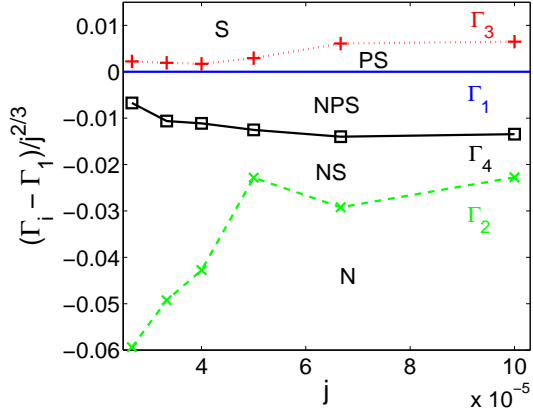}}}
\caption{(Color online) Phase diagram for $u=2.08\approx u_{NS}$ and $j\agt 3j_{\rm co}$. The periodic regime remains stable below $\Gamma_1$, down to the line that we denote by $\Gamma_4$. In the region $\Gamma_4(j) < \Gamma_1(j)$ the three regimes can occur.} 
\label{fig6}
\end{figure}

Let us now examine the phase diagram for $u\agt u_{\rm NS}$. Figure \ref{fig6} shows the diagram for $u\approx u_{\rm NS}$ (in this case we took $u=2.08$). In order to find the P regime we had to look at current densities that are much larger than the current density $j_{\rm co}=7.1\times 10^{-6}$ at which the Hopf singularity occurs, and very close to $\Gamma_1$.  The surprising feature is that the periodic regime leaks beyond $\Gamma_1$, giving rise to a region where all three regimes coexist. We denote by $\Gamma_4(j)$ the new phase separation curve, at which the periodic regime becomes normal, but the stationary state can still persist below it. 
 
\begin{figure}[tbp] 
 \centerline{
    \mbox{\includegraphics[width=5.00in]{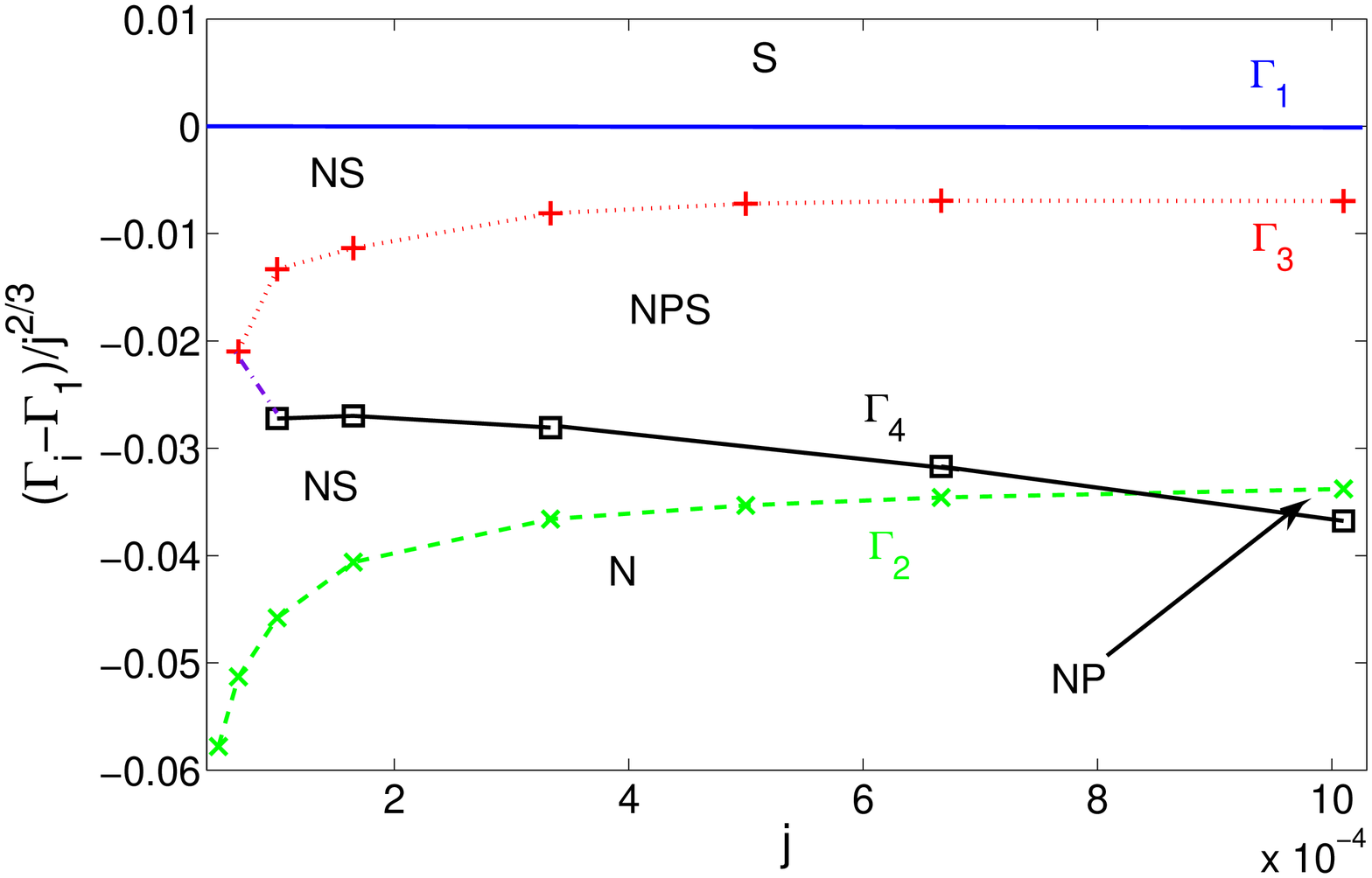}}}
\caption{(Color online) Phase diagram for $u=2.15$ (slightly above $u_{\rm NS}$). The blue line is the stability limit of N, the green line is the stability limit for S, and the region where the periodic regime can exist is bounded between the red line $\Gamma_3$ and the black line $\Gamma_4$. At the red line the periodic phase becomes stationary, and at the black line it becomes normal. The dotted-dashed purple line connects the ends of $\Gamma_3$ and $\Gamma_4$.} 
\label{fig7}
\end{figure}

The migration of the periodic phase to lower $\Gamma$ is seen clearly for $u>u_{\rm NS}$. Figure \ref{fig7} depicts the phase diagram for $u=2.15$. For this value, $\Gamma_1$ separates the S phase from the NS phase, and the whole periodic phase lays underneath a region of NS phase. The red line in Fig.~\ref{fig7} stands for the values at which the periodic phase becomes stationary, and the black line stands for the values at which the periodic phase becomes normal. The small region that has been marked with a purple line will be described elsewhere. We observe that also for this value of $u$ there is a region where all three regimes, N, P and S can occur.

\begin{figure}[tbp] 
 \centerline{
    \mbox{\includegraphics[width=3.00in]{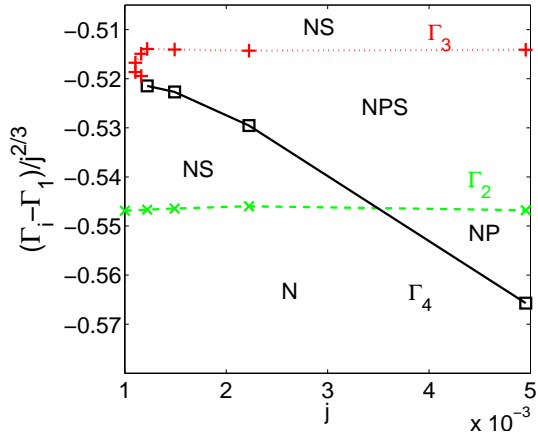}}}
\caption{(Color online) Phase diagram for the realistic value $u=5.79$. As in Fig.~\ref{fig7}, the green line is the stability limit for S, and the region where the periodic regime can exist is bounded between the red and the black lines.} 
\label{physical}
\end{figure}

As $u$ increases further, the periodic phase moves to lower values of $\Gamma$ (relative to $\Gamma_1$) and to higher currents. Figure \ref{physical} shows the phase diagram for the realistic value $u=5.79$. The shape of this diagram is somewhat different than in the case $u=2.15$, but the topology is the same.

We have also investigated the stability limits for $u=5.79$ and small current densities. As in Fig.~\ref{qual}, we found that there is a current density $j=j_{\rm k}<j_{\rm co}$ at which $\Gamma_2$ meets $\Gamma_1$, so that the NS phase exists only if $j>j_{\rm k}$.

\begin{figure}[tbp] 
\centerline{
    \mbox{\includegraphics[width=3.00in]{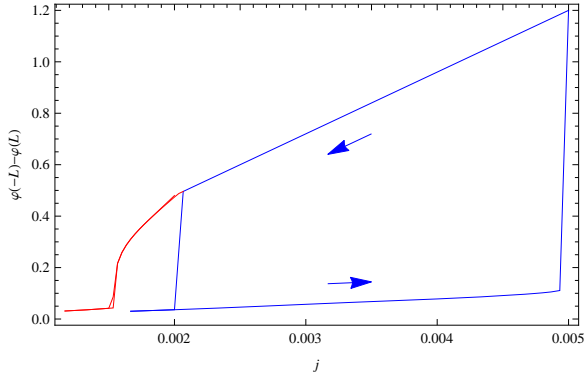}}}
\caption{(Color online) Voltage beteen the extremes of the wire for $u=5.79$ (blue) and for $u=1.67$ (red), as a function of the current density. In both cases $T=0.982T_c$. In the case $u=5.79$ there is a large hysteretic loop. The part of the curve with low voltage corresponds to the stationary regime, and the part with large voltage to the normal regime. In the normal regime the voltage is determined by Ohm's law, and is independent of $u$. In the case $u=1.67$, the periodic regime mediates between N and S, and the slope of the curve changes gradually. The small loop at the heel of this curve is possibly due to slow convergence. } 
\label{Vj}
\end{figure}

The simplest experiment to determine the current pattern in a superconducting wire is to measure the voltage drop $\varphi (-L)-\varphi (L)$ as a function of the current density. Figure \ref{Vj} compares these voltage drops between the cases $u>u_{\rm NS}$ (blue) and $u<u_{\rm NS}$ (red). In the case $u<u_{\rm NS}$ the passage between S and N is mediated by P, and the transition is continuous. On the other hand, for $u>u_{\rm NS}$ there is a region where both N and S are metastable, and the system can remain in its initial regime, until the stability limit is reached. Although, as shown in Fig.~\ref{physical}, the periodic regime could also exist in part of this region, it is never reached by just varying the current. In order to detect the periodic regime, it would be necessary to ``prepare" it, possibly by means of a heat pulse at the middle of the wire.

\section{Conclusions}
The analysis of the possible conduction patterns in a superconducting wire with fixed current is usually considered to be a closed problem. Nevertheless, we have found that it still gives rise to surprises. Rubinstein {\it et al.}\cite{koby} seemingly developed a systematic approach to locate all possible patterns in the parameter plane. We have studied the qualitative changes in the phase diagram when two parameters are varied: the length of the wire and the ratio $u$ between the relaxation times of the order parameter and of the current. Our study was limited to the simplest region in parameter space, where at most one phase slip center is expected.

In the considered region, variation of length does not lead to qualitative modifications, but just to scaling of the phase diagram. On the other hand, variation of $u$ leads to a different topology of the phase diagram. The most remarkable feature is the existence of a region in which any of the regimes (normal, periodic or stationary) is locally stable.

We have estimated the value of $u$ at which the topology of the phase diagram changes.

The conduction pattern can be indirectly inferred by measuring the voltage drop as a function of the current density, as shown in Fig.~\ref{Vj}. For $T_c\sim 10\,$K, $\sigma\sim 10^4\,\Omega^{-1}$cm$^{-1}$ and $\xi (0)\sim 10^{-6}$cm, the units in Eq.~(\ref{units}) are $\varphi_0\sim 1\,$mV and $j_0\sim 10^7$A\,cm$^{-2}$, so that the voltage range in Fig.~\ref{Vj} is of the order of 1\,mV and the current density range is of the order of $10^5$A\,cm$^{-2}$.
Direct observation of the conduction pattern is possible by means of the Dolan--Jackel technique,\cite{DJ} that measures separately the potential $\varphi$ felt by normal electrons and ``the potential felt by the superconducting electrons," $(\arg\psi)_t$, as functions of position. This technique uses local tunnel junctions, through which single electrons or Cooper pairs tunnel to either normal or superconducting probes.

As in Ref.~\onlinecite{koby}, we found that the stationary regime can exist in extensive areas of the temperature--current density plane. Although the stationary regime was already predicted in Ref.~\onlinecite{Bara}, we are not aware of any experimental report of its detection. A possible reason is that most experiments on 1D wires use superconducting rather than normal contacts. Since in the stationary regime $\varphi$ varies with position, whereas $(\arg\psi)_t$ is uniform, the Dolan--Jackel technique could provide convincing evidence for its presence.

\begin{acknowledgments}
This research was supported by the Israel Science Foundation, grant No.\ 249/10. 
\end{acknowledgments}

\end{document}